# Polarization-multiplexing ghost imaging

Shi Dongfeng[1,A)], Zhang Jiamin[1,2], Huang Jian[1,2*], Wang Yingjian[1,2],

Yuan Kee[1], Cao Kaifa[1,3], Xie Chenbo[1], Liu Dong[1] and Zhu Wenyue[1]

[1]*Key Laboratory of Atmospheric Composition and Optical Radiation, Anhui Institute of Optics and Fine Mechanics, Chinese Academy of Sciences, Hefei 230031, China*
[2]*University of Science and Technology of China, Hefei 230026, China*
[3]*School of Environment and Energy Engineering, Anhui Jianzhu University, Hefei 230601, China*
[A]*Email: dfshi@aiofm.ac.cn;\*Corresponding author: jhuang@aiofm.ac.cn*

A novel technique for polarization-multiplexing ghost imaging is proposed to simultaneously obtain multiple polarimetric information by a single detector. Here, polarization-division multiplexing speckles are employed for object illumination. The light reflected from the objects is detected by a single-pixel detector. An iterative reconstruction method is used to restore the fused image containing the different polarimetric information by summing the multiplexed speckles and detected intensities. Next, clear images of the different polarimetric information are recovered by demultiplexing the fused image. The application of this method to the detection of two different polarized objects is presented, and the results clearly demonstrate that the proposed method is effective. An encryption experiment for polarimetric information is also performed by setting the multiplexed speckles' encoding as the keys.

## 1. INTRODUCTION

Ghost imaging (GI) [1-6] employs a non-spatially resolving detector, called as single-pixel detector, to acquire the image of objects. In conventional GI systems, the light is split into two beams using a beam splitter to obtain two correlated light speckles. One beam, which illuminates the objects, is collected by a non-spatially resolving detector, while the other beam is recorded by a spatially resolving detector, e.g., a charged-coupled device (CCD) camera. The signals from the spatially resolving detector don't contain information of the objects. However, the image of the objects can be recovered by calculating the correlation of the signals from the two detectors. Computational GI [1,4,5] uses a computer controlled spatial light modulator (SLM) to remove the beam splitter and spatially resolving detector. Thus, only a non-spatially resolving detector, such as photomultiplier tubes or avalanche photodiodes, is employed as the imaging device in the computational GI system where there is no scanning device. This new imaging system, which includes an integrated computational algorithm, can reduce the cost or size of matrix detectors, especially in the infrared and terahertz regions of the spectrum [2,3], where matrix detectors do not have such good specifications compared to their performance in the visible spectrum.

Polarization [7] is an intrinsic feature of light that provides valuable information of objects beyond that provided by their spectral and intensity distributions. Polarization seeks to measure information of the vector nature of the optical field across a scene. Depolarization is defined as the process of changing polarizated light into unpolarizatd light and reducing the degree of polarization. In fact, when an optical beam interacts with objects, its polarization state almost always changes, and objects of different materials exhibit different depolarization characteristics. Therefore, the light reflected (or transmitted) from the objects encoded by the degree of polarization can be employed to distinguish objects of different materials. It is well known that a promising method for improving the ability of imaging system identification is to employ the polarization components of the reflected light from objects. Thus far, polarimetric imaging with matrix detectors [8-11] has been widely studied and applied in several domains, such as machine vision, biomedical imaging and remote sensing. In GI systems, the influence of light polarization on visibility and signal-to-noise (SNR) has been investigated [12]. The theoretical analysis shows that the visibility of the image of no-depolarization objects increases with the degree of thermal light polarization.

In polarimetric GI (PGI) [13-16] systems, the reflective intensity of objects is decomposed into two different polarization intensities. The reflective and polarimetric information of objects of different materials can be obtained by the PGI system using the correlated method. Computational PGI systems [14,15] with multiple photodetectors to simultaneously obtain images of different linear polarization states have also been proposed. It is also possible to use a single detector to obtain different polarization states in a time-sharing manner [16], but this method does not apply in real-time situations. The results indicate that the connection between polarization and GI can improve the performance of GI in several domains, such as machine vision, biomedical imaging and remote sensing. In this paper, unlike previous studies on PGI, we employ spatial polarization multiplexing to simultaneously acquire multiple polarimetric information by a single detector. An encryption method for polarimetric information is also employed. In the remainder of this paper, we describe the method and present the results.

## 2. THEORETICAL ANALYSIS

The whole procedure of the proposed technique is illustrated in Fig. 1, in which two different polarized objects serve as an example. First, the computer program produces two complementary binary-encoded matrices (B1, B2) and multiplexed speckles (MS). Here, we design the optical path so the two complementary binary-encoded matrices represent two different polarization states. The illumination speckles C are produced by fusing the matrices obtained by multiplying the multiplexed speckles and encoded matrices. The assembling procedure of illumination speckle $C_j$ is shown in the gray frame of Fig. 1. The illumination speckles are projected onto the scene. The reflected light from the two objects passes a polarizer and is detected by a single-pixel detector. The detected values can be expressed as symbol G. The fused image of the two polarized objects can be restored using an iterative reconstruction method. Next, two random sampling images, $f_{1p}$ and $f_{2p}$, indicating different random partial polarimetric information, can be recovered by multiplying the fused image and encoded matrices. Finally, the final images of the different polarimetric information can be obtained using the compressed sensing (CS) algorithm. According to the above analysis, the encoded matrices are very important when recovering accurate information during secondary processing. When the encoded matrices are unknown, accurate polarimetric information cannot be obtained. Accordingly, using the encoded matrices of the proposed method as keys can result in the encryption of the polarimetric information.

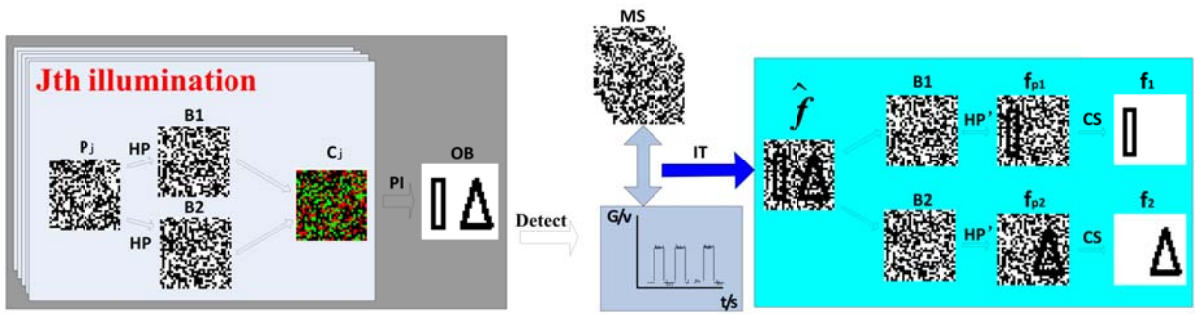

Fig. 1. Procedure of the proposed method: MS: multiplexed speckles, $P_j$: j-th multiplexed speckle; B1 and B2: two complementary binary encoded matrices; $C_j$: j-th illumination speckle where different colors represent different polarization states; HP: Hadamard product of $P_j$ and the encoded matrices; PI: projection illumination; OB: objects; G: intensities detected by the single-pixel detector; IT: iterative reconstruction method; HP': Hadamard product of the fused information $\hat{f}$ and the encoded matrices; $f_{p1}$ and $f_{p2}$: two random partial polarimetric sampling images; CS: compressed sensing algorithm that is used to compute images of different polarimetric information; $f_1$ and $f_2$: two recovered polarimetric images.

Compared with the prior literatures concerning PGI methods, the proposed technique has the following additional advantages. First, multiple polarimetric information can be recovered from a single measurement, which significantly reduces the data acquisition and allows fast data communication to users. Therefore, the system efficiency can be greatly increased using the proposed method. Second, the united process, which incorporates the encryption of the fused image with the compressed acquisition of multiple polarimetric information, ensures the security of private content and thus improves the encryption performance of the PGI system relative to conventional encryption methods [17-20] (in which speckles and/or detected intensities are employed as keys) by providing a third key based on the encoded matrices used in this method.

Our experimental setup is sketched in Fig. 2. Light is split into two beams by a beam splitter (BS). One beam passes through polarizer P1 and becomes horizontal light, is reflected by R1, modulated by the liquid crystal display (LCD1), and then finally enters the polarization beam splitter (PBS). The modulated information of LCD1 can be expressed as the Hadamard product of $P_j$ and binary-encoded matrix B1. The other beam passes through polarizer P2 and becomes vertical light, is reflected by R2, modulated by the liquid crystal display (LCD2), and then finally enters the polarization beam splitter (PBS). The modulated information of LCD2 can be expressed as the Hadamard product of $P_j$ and binary-encoded matrix B2. Polarization-division multiplexing speckles combined by the PBS illuminate the objects. The reflected intensity from the objects (OB) passes through a lens (Le2), linear horizontal polarizer (P3) and filter (F), and then is detected by a single-pixel detector (D).

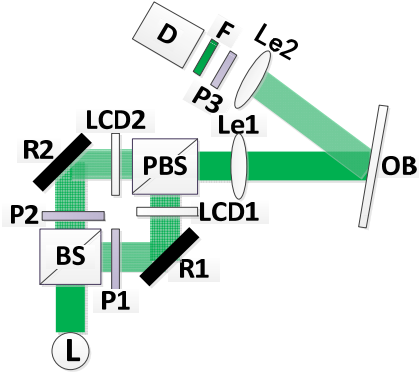

Fig. 2. Configuration of the GI system based on polarization-division multiplexing speckles. L: light source; BS: beam splitter; PBS: polarization beam splitter; P1 and P3: horizontal polarizers; P2: vertical polarizer; R1 and R2: reflected mirrors; OB: objects; Le1 and Le2: lenses; F: filter; D: single-pixel detector; LCD1 and LCD2: liquid crystal displays.

The theory of this method will now be introduced. The Stokes vector $S$[7], which consists of four parameters $(I, Q, U, V)^T$, is employed to describe the polarization state. Here, $I$ represents the total intensity of the light, $Q$ is the difference between horizontal and vertical polarization, $U$ is the difference between linear $+45^0$ and $-45^0$ polarization, and $V$ is the difference between right and left circular polarization. The degree of polarization is a quantity used to describe the portion of an electromagnetic wave that is polarimetric. The polarization state of the reflected light from the objects changes relative to that of the incident light. The relationship between the incident light $S_{in}$ and reflected light $S_{re}$ is given by the following equation:

$$S_{re}(x,y) = M(x,y) S_{in}(x,y), \tag{1}$$

where $M(x,y)$ is the Mueller matrix of the object surface. For light retroreflected off nonbirefringent materials, the Mueller matrix can be approximated as [9,10]

$$M(x,y) = \begin{pmatrix} m_{00}(x,y) & 0 & 0 & 0 \\ 0 & m_{11}(x,y) & 0 & 0 \\ 0 & 0 & m_{11}(x,y) & 0 \\ 0 & 0 & 0 & m_{33}(x,y) \end{pmatrix}, \tag{2}$$

where $m_{00}$ is the reflective coefficient of the objects, and $m_{11}$ and $m_{33}$ are the linear and circular depolarization parameters of the objects, respectively.

Based on Eqs. (1) and (2), we can obtain the reflected light $S_{re}(x,y)=(m_{00}(x,y)*I(x,y), m_{11}(x,y)*Q(x,y), m_{11}(x,y)*U(x,y), m_{33}(x,y)*V(x,y))^T$ whose polarization state can be controlled by the incident light $S_{in}(x,y)=(I(x,y), Q(x,y), U(x,y), V(x,y))^T$. In order to obtain the full Stokes vector $S_{re}(x,y)$, rotating polarizers or multi-detectors are often used. It will either be a slow and storage-heavy task or a complex imaging system. In order to address this problem, the first two components of the Stokes parameters instead of all Stokes parameters are measured; this can also describe well the polarization characteristics of the objects [11]. Existing research methods on PGI [13-16] use multiple detectors or time-difference acquisition to obtain different polarimetric information and recover the object information. In contrast, in this paper, the spatial polarization multiplexing method is employed to acquire multiple polarimetric information simultaneously by a single detector.

Assuming that the objects are sampled at the $j$-th time using the speckle $P_j$, the polarization-division multiplexing speckle has vertical and horizontal polarization states, which is used to illuminate the objects. According to the above analysis, the Stokes vectors of the polarization-division multiplexing speckle can be written as

$$S_{in}(x,y) = P_j(x,y) B_1(x,y) \begin{pmatrix} 1 \\ 1 \\ 0 \\ 0 \end{pmatrix} + P_j(x,y) B_2(x,y) \begin{pmatrix} 1 \\ -1 \\ 0 \\ 0 \end{pmatrix}$$

$$= P_j(x,y) \begin{pmatrix} B_1(x,y) + B_2(x,y) \\ B_1(x,y) - B_2(x,y) \\ 0 \\ 0 \end{pmatrix}, \tag{3}$$

where the first and second terms are the binary light distribution with different polarizations modulated by the two polarizers and LCDs, respectively. Suppose two binary-encoded matrices have the following properties:

$$B_{i1}(x,y)B_{i2}(x,y) = \begin{cases} 0 & i1 \neq i2 \\ B_{i1}(x,y) & i1 = i2 \end{cases},\quad (4)$$

$$\prod B_i = E, \quad (5)$$

where $E$ represents a matrix with all entries are equal to 1, and $\prod$ indicates the accumulation of all matrices. The reflected light from the objects can be expressed as

$$S_{re}(x,y) = P_j(x,y)\begin{pmatrix} m_{00}(x,y)(B_1(x,y)+B_2(x,y)) \\ m_{11}(x,y)(B_1(x,y)-B_2(x,y)) \\ 0 \\ 0 \end{pmatrix}. \quad (6)$$

In front of the detector, a linear horizontal polarizer (P3) is employed to obtain the linearly polarized reflected light intensities. The Mueller matrix of the linear horizontal polarizer can be expressed as [7]

$$M_\parallel = \frac{1}{2}\begin{bmatrix} 1 & 1 & 0 & 0 \\ 1 & 1 & 0 & 0 \\ 0 & 0 & 0 & 0 \\ 0 & 0 & 0 & 0 \end{bmatrix}. \quad (7)$$

Based on Eqs. (1), (6) and (7), the detected intensity from the single-pixel detector can be written as

$$g_j = \frac{1}{2}\sum P_j(x,y)m_{00}(x,y)(B_1(x,y)+B_2(x,y)) \\ + \frac{1}{2}\sum P_j(x,y)m_{11}(x,y)(B_1(x,y)-B_2(x,y)). \quad (8)$$

Further simplifying Eq. (8), we have

$$g_j = \sum_{x,y} P_j(x,y)\hat{f}(x,y), \quad (9)$$

where

$$\hat{f}(x,y) = \frac{1}{2}m_{00}(x,y)(B_1(x,y)+B_2(x,y)) \\ + \frac{1}{2}m_{11}(x,y)(B_1(x,y)-B_2(x,y)), \quad (10)$$

where the fused image $\hat{f}(x,y)$ represents the accumulation of the random samples from multiple polarimetric information. From Eq. (9), we can determine that the acquisition process can be described by the interaction between the multiplexed speckles and fused image. Multiplexed speckles with Hadamard patterns are used in this study. As the illumination speckles produced by the LCD system are binary, we create complementary pairs of illumination speckles $H_+$ and $H_-$ to indirectly access the Hadamard patterns. The intensities detected when illumination speckles $H_{j+}$ and $H_{j-}$ are used to illuminate the objects can be expressed as $g_{j+}$ and $g_{j-}$, respectively, then, the two intensities are subtracted. This process can be written as

$$g_j = g_{j+} - g_{j-} \\ = \sum_{N\times N}(H_{j+}(x,y)\hat{f}(x,y) - H_{j-}(x,y)\hat{f}(x,y)) \\ = \sum_{N\times N}(P_j(x,y)\hat{f}(x,y)), \quad (11)$$

The process, called complementary sensing, has been described in detail [5]. The iterative reconstruction method [4] is employed to recover the fused object information, which can be expressed as

$$\hat{f}(x,y) = \sum_j P_j(x,y)g_j, \quad (12)$$

where $\hat{f}(x, y)$ is the recovered fusion image. Here, the number of multiplexed speckles is equal to $j$. The above formula indicates that the fused image can be expressed as a weighted sum of the multiplexed speckles based on the corresponding coefficients obtained from the detected intensities. The above equation also demonstrates that when the reflected intensities of different polarizations are detected, they are mixed together. Thus, the image containing multiple polarizations is recovered by the iterative reconstruction method. However, we must obtain the image of each individual polarimetric information. Based on the properties of encoded matrices, random samples of each polarization can be obtained as follows:

$$\hat{f}_{p_1} = B_1(x, y)\hat{f}(x, y) \\ = [m_{00}(x, y) + m_{11}(x, y)]B_1(x, y)/2, \tag{13}$$

$$\hat{f}_{p_2} = B_2(x, y)\hat{f}(x, y) \\ = [m_{00}(x, y) - m_{11}(x, y)]B_2(x, y)/2. \tag{14}$$

From the above formula, the Hadamard product of the fused image and encoded matrices can be used to obtain the corresponding random polarimetric information. Based on the polarization character of encoded matrices, $f_{p1}$ and $f_{p2}$ are defined as the horizontal and vertical polarizations, respectively. Next, the complete information of each polarization can be determined with high precision by substituting the encoded matrices and the random sampled image in the CS algorithm. Optimizations can be achieved as follows:

$$f_{pi} = \varphi a_{pi} \quad \text{subject to} \\ \min\left\{\left\|\hat{f}_{pi} - B_{pi}\varphi a_{pi}\right\|_2^2 + \lambda T(a_{pi})\right\}, i = 1 \text{ or } 2 \tag{15}$$

Here, $\varphi$ represents the transformation of the chosen domain resulting in a sparse representation $a_i$, and $\lambda$ and $T$ represent the regularization coefficient and function, respectively. The code employed for CS in this paper [21] is the function *Inpainting_GSR* of the software package. When the encoded matrices are unknown, accurate object information cannot be obtained. Accordingly, the encoded matrices of the proposed method as keys can result in the encryption of the polarization information. According to the information obtained from the two directions of polarization, the intensity and linear polarimetric information can be achieved as:

$$f_I = f_{p1} + f_{p2} = m_{00}(x, y), \tag{16}$$

$$f_P = \frac{|f_{p1} - f_{p2}|}{f_{p1} + f_{p2} + \varepsilon} = \frac{m_{11}(x, y)}{m_{00}(x, y) + \varepsilon}, \tag{17}$$

where $\varepsilon$ is a constant that prevents a division by zero. Next, we will introduce the experimental results.

## 3. EXPERIMENTAL VERIFICATION

The left image in Fig. 3 shows the experimental scene with two piglets. The top and bottom piglets are composed of aluminum and plastic, respectively. The middle and right images in Fig. 3 represent two polarization complementary binary-encoded matrices, where the different polarization states are represented with different colors, respectively. The resolution of the 0.79-inch LCDs is 1024×768 pixels. In the experiments, the central 256×256 pixels of the LCDs with a 2×2 binning model is employed, making the resolution of the object image 128×128. A Thorlabs PMT-PMM02 is used as the single-pixel detector. A central 532nm filter is employed to filter out stray light. The correlation coefficients between the under-sampled images and a reference image are employed to compare the recovered images. This coefficient ranges from zero to one, depending on the resemblance of both images. Our reference image is always the image acquired without under-sampling. The correlation coefficient is calculated with the following function:

$$r = \frac{\sum_m \sum_n (A_{mn} - \overline{A})(B_{mn} - \overline{B})}{\sqrt{\sum_m \sum_n (A_{mn} - \overline{A})^2 \sum_m \sum_n (B_{mn} - \overline{B})^2}}, \tag{18}$$

where $A$ and $B$ are the image matrices with indices $m$ and $n$, respectively, and $\overline{A}$ and $\overline{B}$ represent the mean values of the elements in $A$ and $B$.

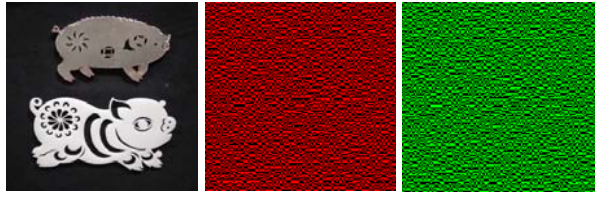

Fig. 3. Experimental illustration: Left is the objects, Middle and Right are the encoded matrices, respectively. The encoded matrices are the 128×128 binary matrices, which contain 8192 elements of 1 and where different colors represent different polarization states. The compression ratio is 50%.

    The evolutionary linear iterative method [4] is employed to recover the fused images in the experiment. The evolutionary linear iteration scheme chooses a subset of Hadamard speckles to recover the fused image of the multiple objects by selecting the patterns with the most significant intensities measured by the single-pixel detector. The recovered fusion images based on the intensities of the reflected light are shown in Fig. 4, and each image presents the results obtained using different numbers of multiplexed speckles. Based on the results, the qualities of the recovered results increase as the number of multiplexed speckles increase. However, we cannot identify the multi-polarization information from the results in Fig. 4. Fortunately, based on the properties of encoded matrices, two images of different polarizations can be obtained by multiplying the fused image with the encoded matrices, as shown in the first and third rows of Fig. 5. Finally, the results of the final recovered images of the horizontal and vertical polarization obtained using CS algorithm are shown in the second and fourth rows of Fig. 5. The experiment was determined to be successful based on the fact that completely accurate image information can be recovered exactly from the compressed samples via the CS algorithm. The correlation coefficients of the horizontal and vertical polarization with the reference information are 0.942, 0.964, 0.989, 0.998 and 0.909, 0.948, 0.988, 0.977 in compression ratios of 12.5%, 25%, 50% and 75%, respectively. The results indicate that the quality of the recovered images is affected by the quality of the fused image and that a positive correlation exists between them. The differences among the images reconstructed from the fused images with the coverage spanning compression ratios from 50% to 100% are almost negligible. According to the results, different encoded matrices can effectively achieve the fusion of multi-polarization information. The compression ratio of the encoded matrices is 50%, so the fusion of two polarimetric information is achieved. If the compression ratio is further reduced, the images of more polarimetric information can be fused and imaged using the results of one measurement. Compared with the traditional PGI system where each of the different polarization states must be individually measured, the proposed method can effectively reduce the data acquisition and improve the imaging efficiency.

    According to the recovered results of Fig. 5, the corresponding intensity and polarimetric images can be obtained according to Eqs. (16) and (17), and the results are presented in Fig. 6. The first row gives the intensity images obtained at different compression ratios and the second row presents the corresponding linear polarimetric images. The results are based on the fact that the depolarization of the metal object is less than that of the plastic object. According to the results shown in the figure, we can use our method to achieve classification of the different polarized objects.

    We use the experimental results to verify the system's encryption capability. In the information encryption transmission process, the detected intensities and multiplexed speckles are transmitted as public information, and only two encoded matrices are transmitted in an encrypted manner. In other words, the encoded matrices are the only keys. It is assumed that, during information transmission, the eavesdropper obtains multiplexed speckles and intensities detected by the single-pixel detector, determines the encoded matrices with a certain level of accuracy, and then uses the partially accurate information to restore the images of the two polarizations. The 100% sampling rate image in Fig. 4 is employed in the experimental study. Fig. 7 shows the results of the restoration of two polarimetric images at different error ratios. The correlation coefficients of the horizontal and vertical polarimetric images with the reference image are 0.711, 0.772, 0.889, 0.957, 0.978 and 0.677, 0.715, 0.823, 0.921, 0.961 in the error rates of 12.5%, 25%, 50%, 75%, 87.5%, respectively. The reference images are acquired without error ratio. It can be seen from the results that the accuracy of the restored polarimetric information increases with the decrease in error ratio. In addition, we note that two phenomena exist; first, because the plastic object exhibits large depolarization characteristics, plastic objects in the two polarization directions have similar information, and the recovered information of such object is negligibly affected by the error ratio; second, because the metal object has a strong ability to protect the polarization, it shows a different phenomenon compared with plastic objects. In other words, the encryption performance of the proposed method is related to the depolarization of the object.

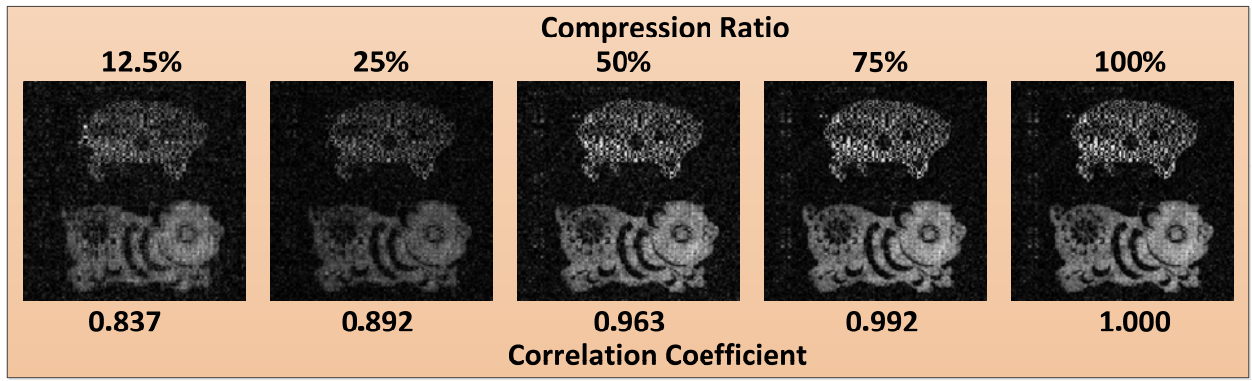

Fig. 4. Fused image reconstruction with different compression ratios. The correlation coefficients between the recovered images with different compression ratios and the reconstruction utilizing a complete Hadamard basis (100% compression ratio) are shown.

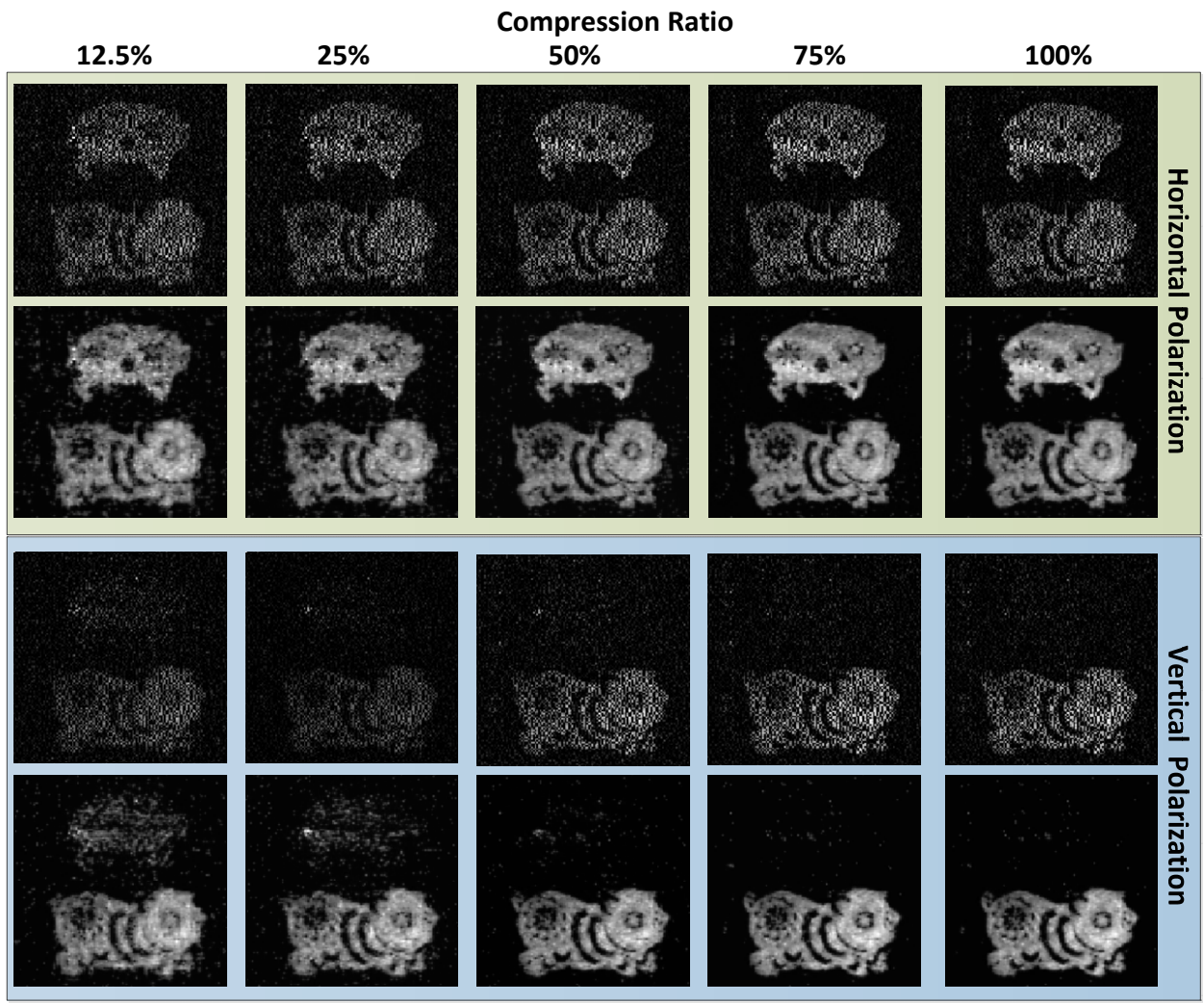

Fig. 5. Multiple polarimetric information reconstructed with different compression ratios using reflected light. The first and third rows present the compressed images obtained by multiplying the fusion result with the encoded matrices, and the second and fourth rows present the final recovered results of the horizontal and vertical polarization.

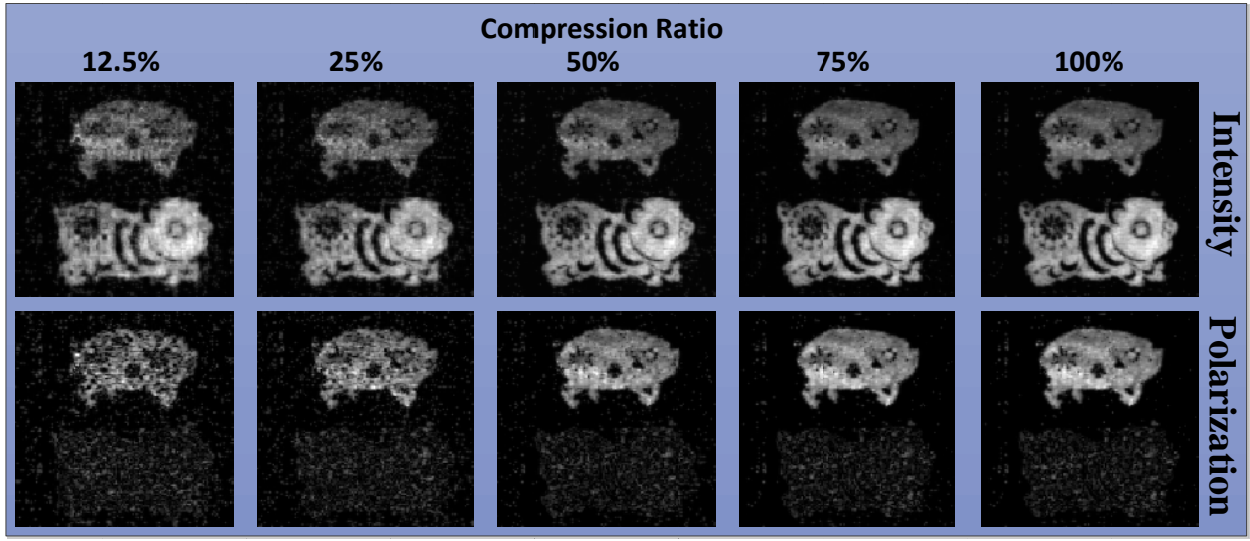

Fig. 6. Recovered results. The first row presents the intensity images with different compression ratios. The second row presents the polarimetric images with different compression ratios.

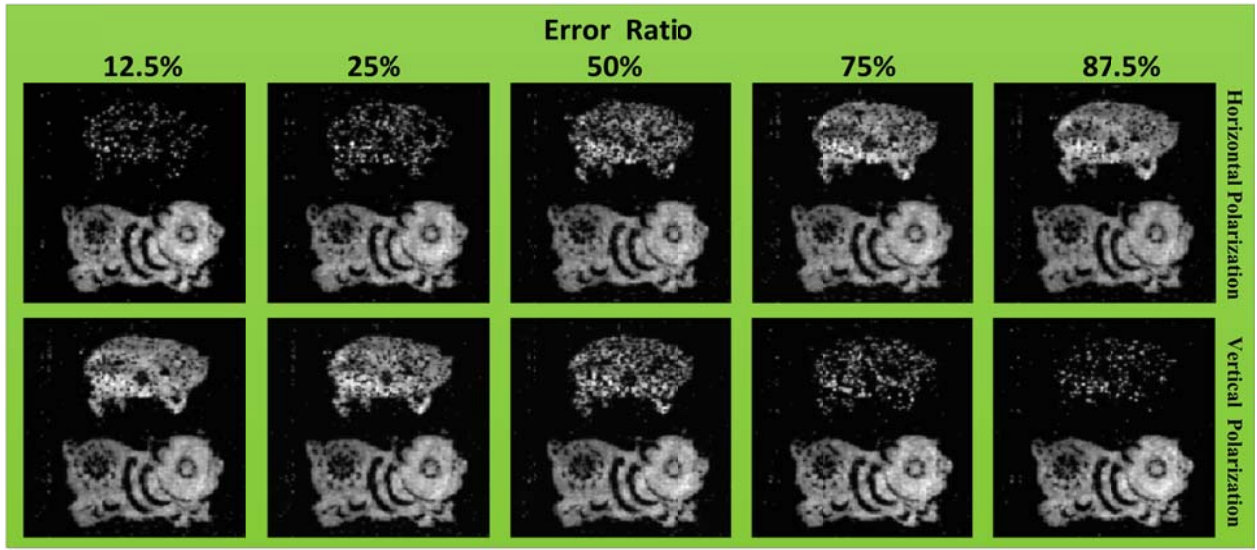

Fig. 7. Image encryption experiment. The results obtained under error ratios of 12.5%, 25%, 50%, 75% and 87.5% for the encoded matrices are shown. The first row presents the images of horizontal polarization with different compression ratios. The second row presents the images of vertical polarization with different compression ratios.

## 4. SUMMARY

In this paper, we first propose polarization-multiplexing GI which is described by theories and experiments. Unlike previous studies, we employ spatial polarization multiplexing to acquire multiple polarimetric information simultaneously by a single detector. In the experiments, the application of this system to the detection of two objects with different materials is presented. It is clearly demonstrated that the proposed method is effective. Of course, a single detector can be employed to obtain more polarimetric information simultaneously by extending this method. For example, three polarimetric information can be achieved simultaneously by a single detector by improving the current maturity of the 3LCD or 3DMD systems. Furthermore, the polarimetric micro-mirror array can be used to directly perform polarization multiplexing modulation of the beam. In general, the central challenge addressed by this method is to find an architecture that effectively balances the final recovery image quality with the number and pattern of the encoded matrices. The choice of encoded matrix plays a key role in image reconstruction. Because various matrices can be chosen, the problem of optimizing the encoded matrix should be carefully studied in the future. Further studies could also involve the sensitivity of the technique to the detection noise, the effects of rough surfaces, the sensitivity of more than two different materials and its performance under atmospheric turbulence.

**Acknowledgements**

This work was supported by the National Natural Science Foundation of China (Nos. 11404344, 41505019 and 41475001) and the CAS Innovation Fund Project (No. CXJJ-17S029).